\documentclass{article}
\usepackage[T1]{fontenc} 
\usepackage[utf8]{inputenc} 
\usepackage{ismir,amsmath,cite,url}
\usepackage{graphicx}
\usepackage{color}
\usepackage{amsfonts}
\usepackage{float}
\usepackage{multirow}
\usepackage[ruled, vlined]{algorithm2e}
\SetKwComment{Comment}{$\triangleright$\ }{}

\usepackage{booktabs}
\usepackage{kotex}
\usepackage{placeins}
\usepackage{amssymb}
\usepackage{pifont}
\usepackage{lineno}

    \title{Polyphonic Piano Transcription Using Autoregressive Multi-State Note Model}

\multauthor
{Taegyun Kwon$^1$ \hspace{1cm} Dasaem Jeong$^{1*}\thanks{* Current affiliation is SK Telecom, South Korea}$  \hspace{1cm} Juhan Nam$^1$} {$^1$ Graduate School of Culture Technology, KAIST, South Korea\\
{\tt\small \{ilcobo2, jdasam, juhan.nam\}@kaist.ac.kr}
}
\def\authorname{Taegyun Kwon, Dasaem Jeong, Juhan Nam}

\begin{document}

\maketitle
\begin{abstract}
Recent advances in polyphonic piano transcription have been made primarily by a deliberate design of neural network architectures that detect different note states such as onset or sustain and model the temporal evolution of the states. The majority of them, however, use separate neural networks for each note state, thereby optimizing multiple loss functions, and also they handle the temporal evolution of note states by abstract connections between the state-wise neural networks or using a post-processing module. In this paper, we propose a unified neural network architecture where multiple note states are predicted as a softmax output with a single loss function and the temporal order is learned by an auto-regressive connection within the single neural network. This compact model allows to increase note states without architectural complexity. Using the MAESTRO dataset, we examine various combinations of multiple note states including on, onset, sustain, re-onset, offset, and off. We also show that the autoregressive module effectively learns inter-state dependency of notes. Finally, we show that our proposed model achieves performance comparable to state-of-the-arts with fewer parameters. 

\end{abstract}
\section{Introduction}\label{sec:introduction}

Automatic music transcription (AMT) refers to an automated process that converts musical signals into a piano roll or a musical score. Polyphonic piano transcription is a specific AMT task for piano music. Due to the complex nature of piano sound such as overlapping spectra and interference among notes and inharmonic overtones, most of recent approaches are based on learning algorithms such as non-negative matrix factorization (NMF) and deep neural networks (DNN)~\cite{Benetos2019}. In particular, the transcription performance has been significantly improved by virtue of the representation power of DNN~\cite{sigtia2016end,onpotential,hawthorne2017onsets,kimadversarial} and large-scale piano music data such as the MAESTRO dataset~\cite{maestro}. 

A key element in designing state-of-the-art DNN architectures is detecting multiple states of a note beyond the conventional binary states (i.e., on/off) and modeling the temporal evolution of the note states. For example, the \textit{Onsets and Frames} model incorporated a note onset detection network into the frame-level pitch detection network \cite{hawthorne2017onsets}. This onset-aware model significantly reduced note-level false positive errors, which is critical in perceptual evaluation of the transcription. Similar multi-state note modeling approaches are found in \cite{duan2014note,Cogliati2015,kwon2017,hawthorne2017onsets,Two-Stage} and some detect even more phases of note envelope including onset, sustain and offset \cite{Costantini2009,kelz2019adsr}. As such, various versions of note state representations have been suggested so far and showed improved performances. However, the majority of them use separate neural networks for each note state. This requires to optimize multiple loss functions, progressively increasing the model complexity for more note states. Also, they handle the temporal evolution of note states by an abstract connection between the hidden layers of the state-wise neural networks \cite{hawthorne2017onsets} or using a separate neural network to model piano-roll data~\cite{ModelingHigh2012,sigtia2016end,wang2018polyphonic}. 

In this paper, we propose a unified neural network architecture where individual neural networks for each state and the temporal order modeling are integrated within a single neural network. We implement the all-in-one architecture by predicting multiple states of a note as a softmax output and modeling the temporal order in an auto-regressive connection in the output layer. Specifically, the architecture is composed of convolution neural network (CNN) and recurrent neural network (RNN). For each time step, the CNN module summarizes local acoustic features into a frame-level latent vector. The RNN modules predicts the note states from the softmax outputs conditioned on the latent vector and the previous outputs via an auto-regressive connection. The multi-class classification approach for note states allows the model to increase the number of note states without architectural complexity. Taking the advantage of this property, we examine various combinations of multiple states including \textit{on}, \textit{onset}, \textit{sustain}, \textit{re-onset}, \textit{offset}, and \textit{off} by comparing the performances and visualizing an example of the note state activations. In addition, we show that the autoregressive module effectively learns inter-state dependency by comparing it to a non-auto-regressive version. Finally, we show that the auto-regressive multi-state note model can achieve transcription performance comparable to the state-of-the-art \textit{Onsets and Frames} model on the MAESTRO dataset.

\section{Related Works}
\label{sec:format}
\subsection{Multi-State Note Modeling}

Most of recent approaches in polyphonic piano transcription are based on deep learning. The model architectures are diverse, including CNN~\cite{sigtia2016end,onpotential,Two-Stage,kelz2019adsr}, RNN~\cite{rnn_transcription,kwon2017}, CRNN~\cite{hawthorne2017onsets,kimadversarial,Carvalho2017}, and U-Net\cite{kelz2017experimental}. The loss function is typically the cross-entropy between predicted and ground truth labels but also includes the adversarial loss \cite{kimadversarial}. An important direction in designing a neural network architecture is detecting note onset explicitly apart from the binary on/off states~\cite{Two-Stage,hawthorne2017onsets,kwon2017,kelz2019adsr}, considering that piano sound starts with a percussive tone but, after the attack park, it slowly decays with a harmonic tone \cite{benetos2013automatic}. This multi-state note modeling even including note offset was already explored before the DNN approaches become dominant~\cite{Costantini2009,duan2014note,Cogliati2015}. While this multi-state note modeling has significantly improved the transcription performance, to the best of our knowledge, there has been no study that systemically compares various combinations of multi-state note representations. In this paper, we define five note states even considering the sustain pedal that globally changes the note states, and examine the different representations of temporal evolution. 

\subsection{Temporal Modeling of Multi-State Notes}

Modeling the temporal order of note states is an essential step to improve the transcription performance~\cite{Benetos2019}. A popular choice is hidden Markov model (HMM), which learns the temporal dependency of note states typically for each pitch. The note states can range from binary (on/off) \cite{poliner2006discriminative,nam2011,Cazau2017ImprovingNS} to more complete note phases (attack, decay, sustain, and release) \cite{kelz2019adsr}. Another approach is the autoregressive modeling which has been implemented with an RNN or its variants~\cite{ModelingHigh2012,sigtia2016end,wang2018polyphonic}. The autoregressive models can learn much wider musical context than HMM, covering inter-note and long-term dependency at the cost of sophisticated decoding algorithm~\cite{ModelingHigh2012}. Since this is analogous to the language model in speech recognition, it is also called musical language model (MLM). The MLMs are trained only with frame labels without paired audio data. This enables them to leverage large-scale symbolic data such as MIDI files. However, this decoupling from audio data may not take advantage of the synergy when the MLM is conditioned with the acoustic information, for example, using a  transduction model~\cite{boulanger2013transduction,Ycart2018PolyphonicMS}. The \textit{Onsets and Frames} model learns the temporal order of note states without an MLM by having a directed connection between different columns of neural networks that account for onsets and frames states, respectively~\cite{hawthorne2017onsets}. However, this hidden-layer connection implicitly learns the temporal orders and the design choice is heuristic.

Our proposed architecture integrates the acoustic model with the MLM by conditioning the autoregressive multi-state note modeling on the acoustic latent features. Unlike the transduction models that use the pre-trained features or posterior as input~\cite{boulanger2013transduction,Ycart2018PolyphonicMS}, we train the acoustic model and MLM jointly within a single neural network in an end-to-end manner. This unified approach was recently attempted by the image-to-image translation model between mel-spectrogram and piano-roll \cite{kimadversarial}. However, our model has multiple states for each note and the state transitions are learned via the autoregressive connection.

\section{Proposed Method}

\subsection{Term Definitions}
\label{formation}
In this paper, we will use frame as an unit of time. $t$ indicates the frame index.
For a given audio recording, we express its audio feature as $\mathbb{X}=\{x_t\}$, where $x_t$ is a vector that represents local audio feature at $t$, and $\mathbb{N}=\{n_i\}$, where $n_i$ denotes a musical note with index $i$ over the frames. 
Each note consists of onset, offset time and pitch. We also represent notes with a piano-roll-like form $\mathbb{Y} = \{y_t^p\}$, where $y_t^p$ denotes a frame-level state of a note with pitch $p$ at time step $t$. The pitch $p$ corresponds to each key of piano. 
The actual form of $y_t^p$ are determined by the chosen note state representation and network architecture. 
For example, the \textit{Onsets and Frames} model 
uses three parallel binary rolls $\mathbb{Y}_{onset}$, $\mathbb{Y}_{frame}$, and $\mathbb{Y}_{offset}$, as they have three columns of networks for separate detection of onset, offset and frame. 
In our proposed model that uses a single-column network, $y_t^p$ is a one-hot vector of multiple states. Without the pitch superscript, $y_t$ denotes concatenated one-hot vectors ($y_t^p$) for all pitches (88 keys in piano) at time step $t$. In Section \ref{sec:note_states}, we explain various combinations of multiple states of a note. 


\subsection{Model Overview}
\begin{figure}
 \includegraphics[width=\columnwidth]{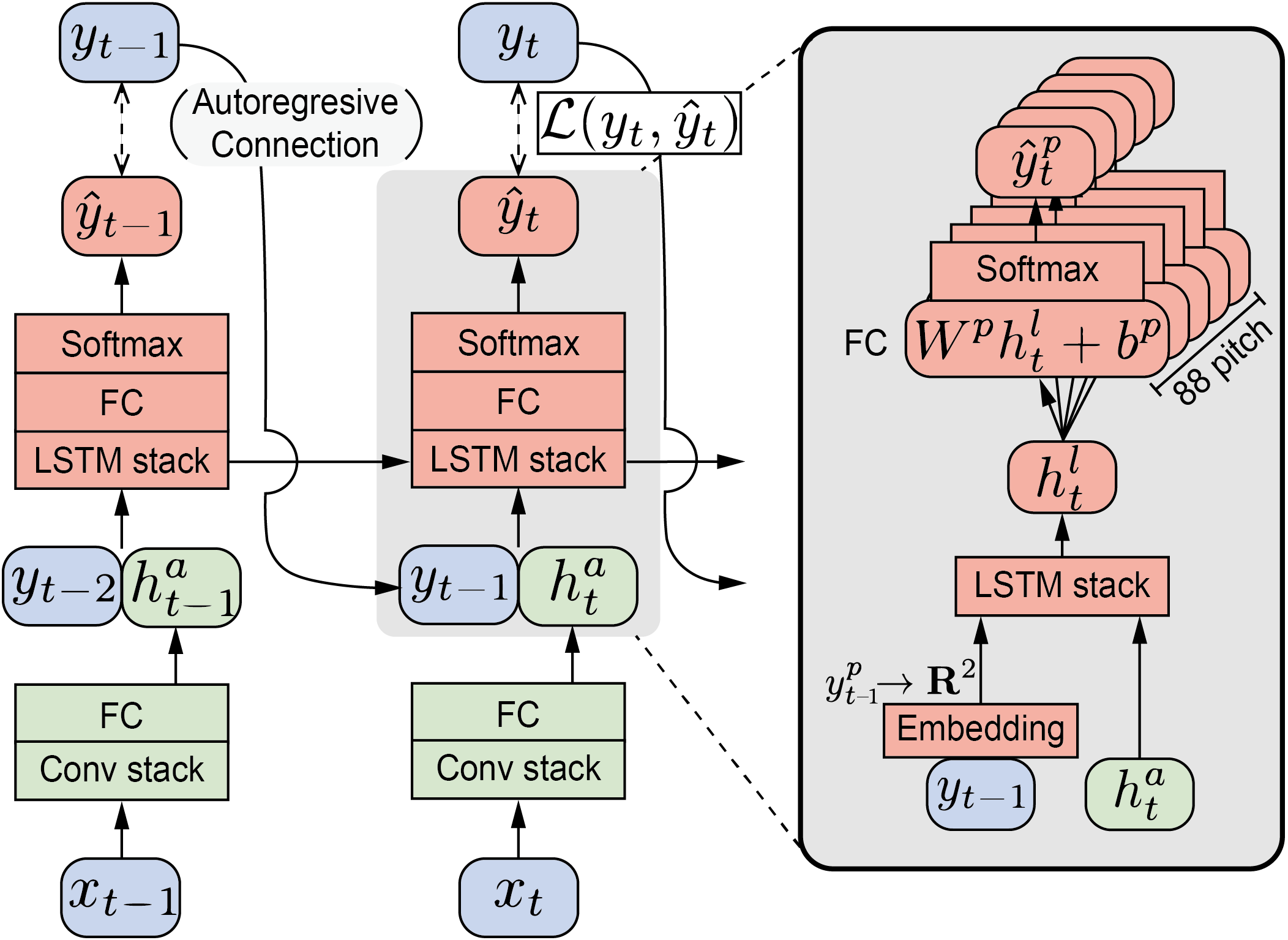}
 \caption{A diagram of the proposed CRNN architecture at time step $t$.}
 \label{fig:model_diagram}
\end{figure}

Our proposed network architecture consists of two modules as shown in \figref{fig:model_diagram}.
The first module is a CNN-based acoustic model to extract local feature $h^a_t$ from the input $x_t$. The second module is an RNN-based autoregressive MLM to estimate the output $y^t$ from the previous output $y^{t-1}$ but it is conditioned on the extracted audio feature $h^a_t$. The output layer have 88 independent softmax functions, each of which corresponds to pitch $p$. The following equations summarize the input and output in each module. 

\begin{equation}
\begin{split}
h^a_t &= CNN(x_t)\\
h^l_t &= RNN(y_{t-1}, h^a_t) \\
\hat{y}^p_t &= softmax(W^p h^l_t + b^p)
\end{split}
\end{equation}

For the acoustic model, we borrow the CNN architecture (\textit{convnet}) proposed in \cite{onpotential}, which is also used in \cite{hawthorne2017onsets, kimadversarial}. The acoustic model consists of three convolutional layers followed by a fully-connected layer. For the autoregressive MLM, we employ two layers of uni-directional long short-term memory (LSTM). 
When the one-hot vector $y^p_{t-1}$ at the previous time step $t\mkern1.5mu{-}\mkern1.5mu1$ is used as the input of the LSTM stack, it is embedded into a two-dimensional vector with continuous values to be matched with the audio feature $h_t^a$, another input of the LSTM stack. 
Since all of note states are predicted from the single CRNN architecture, our model has fewer parameters compare to the \textit{Onsets and Frames} model. In addition, the causal uni-directional RNN in our model enables real-time applications.

\begin{figure}[t]
 \includegraphics[width=\columnwidth]{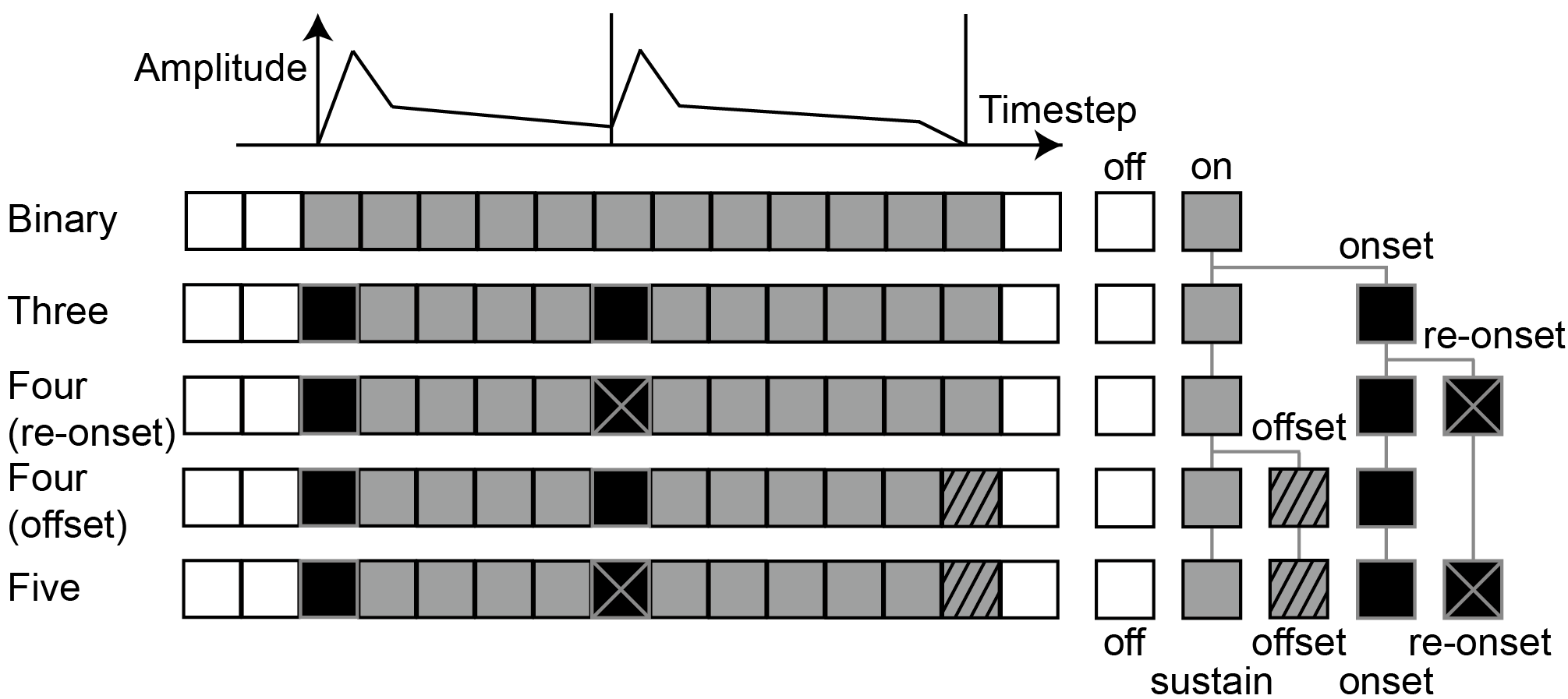}
 \caption{Visualization of five note state representations. The idealized ADSR curve of two consecutive notes and corresponding annotations are displayed.}
 \label{fig:note_states}
\end{figure}

\subsection{Multiple Note States}
\label{sec:note_states}
We illustrate five different note state representations in \figref{fig:note_states}. Our main idea is to extend the conventional binary state (\textit{onset} and \textit{off}) to multiple states using the states transition of note activations such as note \textit{onset} and note \textit{offset}. Considering that sustain pedal affects the note transition, we also add \textit{re-onset}, the moment that a new note is played while the previously played note on the same key is being sustained with the pedal. In addition to the transition states, we distinguished \textit{sustain} from \textit{on}. Using the multiple note states, we examine five types of note state representations: binary, three states (\textit{off, onset, sustain}), four states which have additional \{\textit{re-onset} or \textit{offset}\} state, and five states which utilize all local states.



We define the note state representation over the softmax output in the LSTM in contrast to other multi-note-state models which have a separate binary classifier for each state~\cite{hawthorne2017onsets,kelz2019adsr}. 
\begin{equation}
\mathcal{L}(y, \hat{y}) = -\sum_{t=0}^{T} \sum_{p=p_{min}}^{p_{max}} \sum_{i=0}^{\#states} y^p_t(i) log (\hat{y}^p_t(i))
\end{equation}
We expect two advantages from this multi-class approach. When the states are explicitly defined by a single variable, the relation among states becomes concise. This may help the autoregressive model to learn note transition patterns more easily. Also, it prevents unrealistic combination of states which can be occur in inference. (i.e. $P(onset)\mkern1.5mu{=}\mkern1.5mu1$ but $P(on)\mkern1.5mu{=}\mkern1.5mu0$). Adding \textit{re-onset} class independently, out of onset or sustain class, is also expected to be helpful to train the neural network due to the distinct percussive timbre of piano at note attack.

\subsection{Autoregressive Model}
For given a audio recording, our goal is estimate the notes $\mathbb{N}$ from the acoustic features $\mathbb{X}$.
In practice, it is common to estimate the frame-level note states first  $\mathbb{Y}$ by computing the maximum-likelihood of  $P(\mathbb{Y}|\mathbb{X})$ and then to decode the estimated $\mathbb{Y}$ into $\mathbb{N}$. 
In the majority of AMT algorithms, the condition probability is factorized as follows (see \cite{sigtia2016end } for details):
\begin{equation}
\label{eq:factorize}
P(\mathbb{Y}|\mathbb{X}) \propto P(y_0|x_0)\prod_{t=1}^{T} P(y_t|y_0 ... y_{t-1})P(y_t|x_t)
\end{equation}
where $P(y_t|x_t)$ corresponds to an acoustic model and $P(y_t|y_0 ... y_{t-1})$ accounts for a (musical) language model. 
The factorization allows the MLM to be trained with large-scale symbolic data such as MIDI without paired audio recordings. However, this approach has two issues. First, the factorization may not take advantage of the synergy when the MLM is conditioned with the acoustic information as input, for example, using a transduction model~\cite{boulanger2013transduction,Ycart2018PolyphonicMS}. Second, the frame-level MLM in Equation \ref{eq:factorize} is usually set up to learn the dependency of binary on or off states over piano-rolls \cite{ModelingHigh2012,sigtia2016end}. The recurring nature of the binary representation may lead the model to play a role of smoothing, rather than learning any kind of musical
structure \cite{sigtia2016end,ycart2017study}. While a note-level MLM (i.e, $P(n_i|n_{\leq i-1})$) can solve this problem \cite{ycart2017study, wang2018polyphonic} (and also this learns the distribution of notes more meaningfully as notes in AMT are analogous to words in speech recognition), it requires a separate beat and meter detection algorithm to convert the time unit \cite{Ycart2019BLENDINGAA}. 

Our proposed method addresses the two issues by 1) jointly training the acoustic model and MLM and 2) using multiple note states. Unlike the transduction models that use the pre-trained features or posterior as input~\cite{boulanger2013transduction,Ycart2018PolyphonicMS}, we train the acoustic model and MLM in an end-to-end manner through the auto-regressive CRNN architecture. More precisely, we express the condition probability by conditioning the autogressive model on the acoustic input.   

\begin{equation}
P(\mathbb{Y}|\mathbb{X}) \propto P(y_0| x_0) \prod_{t=1}^{T}P(y_t|y_{\leq t-1}, x_{\leq t})
\end{equation}
Although our model maintains the frame-level MLM, the note-aware multiple state representation may mitigate the repeated patterns of the simple binary representation.

\subsection{Note Decoding}
\label{sec:decoding}


Once we estimates frame-level note states, we decode them into musical notes in the inference phase. We examine two ways of note decoding. One is a simple greedy approach and the other is a global optimization strategy using a modified beam search. The simple greedy decoding samples from the most probable estimated state of  $\hat{y}_t$  every frame. However, it does not guarantee global optimum over multiple time steps. On the other hand, it is intractable to examine all possible sequences especially in the high-dimensional sequence. To overcome this problem, a beam search is usually used to obtain a global optimum. 
For MLM, a high-dimensional beam search\cite{boulanger2013transduction} or a hashed beam search\cite{sigtia2016end} was proposed to reduce the complexity in high dimensional situation. However, their methods mainly aim to capture dependency across pitches on the binary piano-roll representation. To focus on dependency over the multiple states, we simplified the beam search to find the optimum for every pitch independently. 
We achieve this by examining the sub-sequences in the beam search tree only when the second-best state of a pitch has a higher possibility than a certain threshold. When such pitches and frames are detected, we perform beam-search for five more steps only with that pitch while fixing other pitches with greedy sampling. We also consider only two states with the highest probability at each frame because we expect the probability becomes negligible from the third in most cases. After the best path for the pitch is found, we proceeded to other pitches and next frame recursively. Most of the target frames were close to onset and offset. Therefore, the pitch-wise path search can be regarded as locating the onset or offset to an optimal position. This method operates like the greedy decoding if all of the estimated frames have a high confidence.

After note states are inferred by one of the decoding algorithms, we apply a simple rule to determine notes.  
Along with frame axis, a note is initiated if a \{\textit{onset} or \textit{on}\} state is detected after \textit{off}. The offset of the note is determined if \{\textit{offset} or \textit{off}\} is detected after the note initialization.



    


\begin{table*}[!t]
\centering
\resizebox{\textwidth}{!}{%
\begin{tabular}{@{} {c}*{15}{c} @{}}
\toprule
\multirow{2}{*}[-0.5em]{Models} &\multirow{2}{*}[-0.5em]{AR} &\multicolumn{3}{c}{Used states} &\multirow{2}{*}[-0.5em]{\#Parameters} & \multicolumn{3}{c}{Frame} & \multicolumn{3}{c}{Note Onset} & \multicolumn{3}{c}{Note with Offset} \\ \cmidrule(lr){3-5}  \cmidrule(lr){7-9} \cmidrule(lr){10-12} \cmidrule(lr){13-15}
    &     & \textit{onset} & \textit{re-onset} & \textit{offset} & & precision            & recall            & F-score          & precision            & recall            & F-score       & precision            & recall            & F-score       \\ \midrule
    Binary  & $\bigcirc$ & & &  & 14.4M & 0.7768       & 0.8881       & 0.8161      & 0.9874   & 0.6281   & 0.7560   & 0.6399    & 0.4278   & 0.5064     \\ 
Binary &   & & & & 13.9M  & 0.9128       & 0.8815       & 0.8961      & 0.8184   & 0.6646   & 0.7279   & 0.5655    & 0.4696   & 0.5095 \\ \midrule
Three & $\bigcirc$  & $\bigcirc$ & & & 14.5M & 0.7643  & 0.8828  & 0.8047 & 0.9878  & 0.9042 & 0.9433 & 0.7949 & 0.7249  & 0.7593        \\ 
Three &    & $\bigcirc$ & & & 13.9M & 0.9502       & 0.8104       & 0.8740      & 0.9907   & 0.8247   & 0.8985   & 0.8259    & 0.6904   & 0.7508   \\
 \midrule
Four(offset)  & $\bigcirc$  & $\bigcirc$ & & $\bigcirc$ & 14.5M & 0.7957       & 0.8774       & 0.8258      & 0.9874   & 0.9015   & 0.9416   & 0.8105    & 0.7409   & 0.7735   \\ 
Four (offset)  &   & $\bigcirc$ & & $\bigcirc$ & 14.0M &  0.9549       & 0.8080       & 0.8745      & 0.9890   & 0.8392   & 0.9065   & 0.8264    & 0.7039   & 0.7590   \\ \midrule
Four (re-onset)  & $\bigcirc$ & $\bigcirc$ & $\bigcirc$ &  & 14.5M & 0.7834       & 0.8864      & 0.8207      & 0.9871   & 0.9103   & 0.9465   & 0.8074    & 0.7450   & 0.7744   \\ 
Four (re-onset)  &  & $\bigcirc$ & $\bigcirc$ & & 14.0M & 0.9529       & 0.8049       & 0.8717      & 0.9928   & 0.8258   & 0.8997   & 0.8313    & 0.6946   & 0.7553   \\ \midrule
Five  & $\bigcirc$  & $\bigcirc$ & $\bigcirc$ & $\bigcirc$   & 14.6M  & 0.8191       & 0.8732       & 0.8382      & 0.9856   & 0.9121   & 0.9467   & 0.8259    & 0.7648   & 0.7936   \\ 
Five  &    & $\bigcirc$ & $\bigcirc$ & $\bigcirc$   & 14.1M   & 0.9391       & 0.8170       & 0.8730      & 0.9923   & 0.8278   & 0.9009   & 0.8213    & 0.6878   & 0.7472   \\ \midrule[0.1em]
Five (small)  & $\bigcirc$  & $\bigcirc$& $\bigcirc$ & $\bigcirc$  & 6.1M  & 0.7264 & 0.8933   & 0.7889      & 0.9889   & 0.9043   & 0.9438   & 0.7755    & 0.7093   & 0.7403   \\ \midrule[0.1em]
\multicolumn{5}{l}{Onsets and Frames (paper)\cite{maestro}} & 18.3M$*$ & 0.9211       & 0.8841       & 0.9015      & 0.9827   & 0.9261   & 0.9532   & 0.8295    & 0.7824   & 0.8050 \\ 
\multicolumn{5}{l}{Onsets and Frames (pretrained)$^1$} & 23.5M & 0.8737       & 0.8768       & 0.8733      & 0.9792   & 0.9182   & 0.9473   & 0.8114    & 0.7615   & 0.7853   \\ 
\multicolumn{5}{l}{Onsets and Frames (reimplemented) } & 18.3M & 0.9350       & 0.8771       & 0.9045      & 0.9939   & 0.8993   & 0.9436   & 0.8135    & 0.7371   & 0.7730 \\ 
\multicolumn{5}{l}{Onsets and Frames Uni-LSTM (reimplemented)} & 15.6M & 0.9356       & 0.8599       & 0.8954      & 0.9929   & 0.8917   & 0.9388   & 0.8028    & 0.7218   & 0.7595 \\ 
\multicolumn{5}{l}{Non-Saturating GAN (paper)\cite{kimadversarial}} & 26.9M$*\dag$ & 0.931       & 0.898\        & \textbf{0.914}      & 0.981   & 0.932   & 0.956   & 0.835    & 0.793   & 0.813 \\ 

\bottomrule
\end{tabular}
}
\caption{Frame and note metrics for the five note state representations. All measures are based on decoded sequence with greedy decoding. Precision, Recall and F1 score are averaged over piece-wise results. AR stands for `autoregressive'. Refer \ref{sec:hparam} for detail. * This number was estimated based on hyperparameters in the paper. \dag This includes a neural network to estimate note velocity. }
\label{tab:five_state_metric}
\end{table*}

\section{Experiments}

\label{sec:experiments}
\subsection{Dataset}
We trained our model with the MAESTRO dataset v1.0.0 \cite{maestro} with the published training and test split. 
The dataset provides 1184 performances played in the \textit{International Piano-e-Competition}. Both audio recordings and the corresponding aligned MIDI files captured through Disklavier are given. To compensate the effect of sustain pedal on note offsets in labeling, we elongated the offset of notes according to the sustain pedal followed by methodology in \cite{hawthorne2017onsets}.


\subsection{Metrics}
We employed the standard frame-based and note-based metric using \textit{mir\_eval} \cite{mir_eval} package. We report precision, recall and F1 score. We used a threshold of $50msec$ to detect note onset. We counted note offset as hit when the difference is within $50msec$ or $\pm 20\%$ of note duration. 

\subsection{Hyperparameters}
\label{sec:hparam}
We used log-compressed mel-spectrograms as input of the acoustic model. We computed the mel-spectrograms with 229 logarithmically-spaced frequency bins, a hop length of 512, an FFT window of 2048, and a sample rate of 16kHz following \cite{hawthorne2017onsets,kimadversarial}. The CNN consists of $48/48/96$ nodes from the bottom to the top layer, and the following fully-connected (FC) layer has 768 nodes. The RNN consists of two layers of LSTM with 768 nodes. The output FC layer has $(88\mkern1.5mu{\times}\mkern1.5mu N)$ nodes where $N$ is the number of note states. The receptive field size of the top hidden layer in the CNN is 7 frames of mel-spectrograms centered at time step $t$. This yields 176 msec latency when the model runs in real-time.  


We used the categorical cross entropy loss and Adam optimizer for training, applying a batch size of 32 and a learning rate of $\mkern1.5mu6e\mkern1.5mu{-}\mkern1.5mu4 $ with a decay of 0.02 in every 10000 steps. We trained our model with the teacher-forcing method, which provides ground-truth data of the previous time step for the AR layer. We tried scheduled sampling \cite{bengio2015scheduled} but our preliminary result showed a significant degradation in performance. We randomly segmented audio into $10sec$ while training, but the whole sequence is transcribed at once during inference. We evaluated the models after 200k steps in training. In addition to the aforementioned hyperparameters, we report the experimental result of a smaller model where the number of nodes in the LSTM is set to 256 (three times smaller than the above) to reduce the number of model parameters.

\subsection{Comparative Experiments}
We conducted a comparative experiment with the proposed method with the five note state representations. Also, we evaluated the same set of models without the autoregressive connection to verify the effectiveness of the autoregressive model. 
For the autoregressive models, we evaluated the note decoding results as well. We compared our model mainly with the \textit{Onsets and Frames} model and its GAN-based extension\cite{kimadversarial}, which is also trained on the same dataset. In addition to the reported performance, we evaluate \textit{Onsets and Frames} model with the publicly available pre-trained model\footnote{https://github.com/tensorflow/magenta, accessed on Sep 14, 2019} and our own re-implementation. 

\section{Results}
\label{sec:result}


\subsection{Effect of Multi-State Note Representations }
We report the averaged metrics over the test set in \tablename~\ref{tab:five_state_metric}. Overall, the use of \textit{onset} makes a significant improvement as seen in comparison between the `Binary' model and other multiple models. 
This result is in accordance with previous studies \cite{hawthorne2017onsets}. The note-with-offset score also increases along but this might be seen as affected by the increased number of matched notes. The `Binary' model achieves a high frame-level F1-score but at the same time it has the lowest note onset F1-score. We suspect that overlapped notes without offsets (notes with \textit{re-onset}) were not distinguishable in the binary note state representation, thereby leading low recall in the note onset score.


The influence of \textit{re-onset} and \textit{offset} is observed from the note-with-offset score. The `Four' models that have either one of the states achieve higher scores than the `Three' model. The `Five' model that has both states achieves a higher score than the `Four' models, particularly with the AR connection. However, \textit{re-onset} and \textit{offset} do not help improving the note onset score much. 


Among the five note state representations, the `Five' model achieves slightly higher frame-level and note-level F-scores than others. We investigated the model further by downsizing the LSTM units (small). The small model has a lower F-score in the note-with-offset score but it achieves comparable F-scores to the original `Five' model in the note onset score.

\begin{figure}[!t]
\includegraphics[width=\columnwidth]{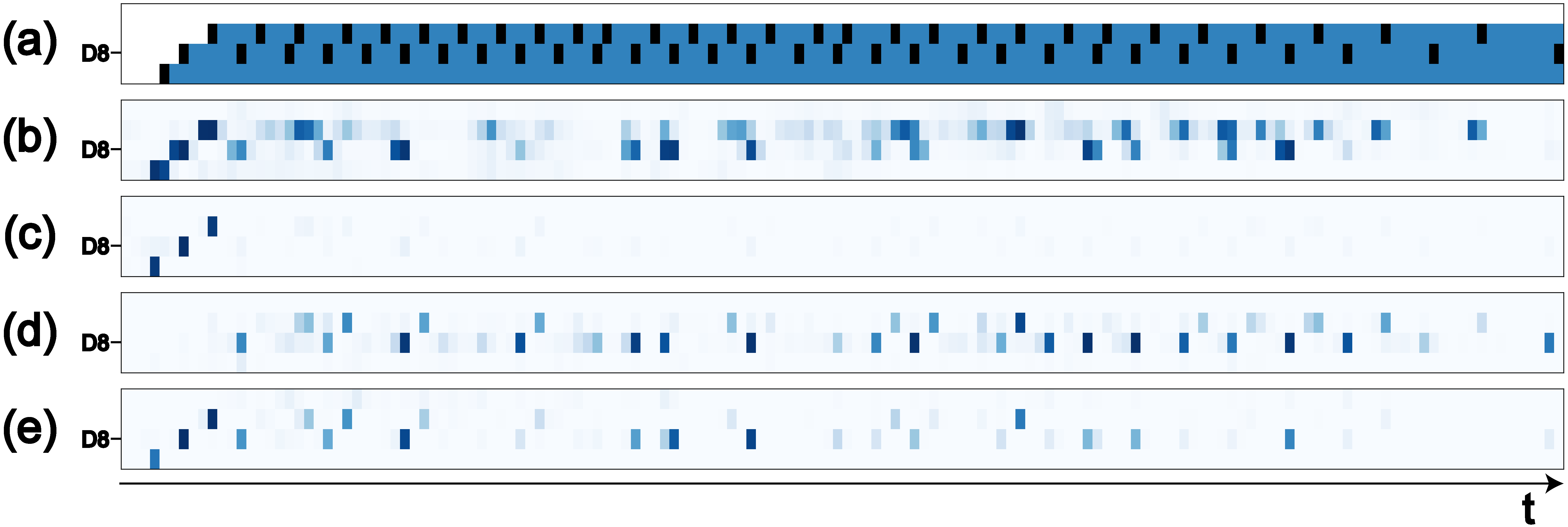}
    \caption{Comparison between \textit{onset} and \textit{re-onset} activation on trill notes. (a) ground truth (b) onset activation of \textit{Onsets and Frames}   (c) \textit{onset} activation of `Five' AR model (d)  \textit{re-onset} activation of `Five' AR model (e) \textit{onset} activation of `Four (offset)' AR model}
\label{fig:reonset}
\end{figure}

We also observed that \textit{re-onset} in the `Five' model estimates sharper activations compared to the common \textit{onset} when it estimates repeated notes with extremely short intervals such as trills, as shown in \figref{fig:reonset}. The number of such note patterns is too small to affect overall performance but it would be helpful when detailed analysis on articulation is necessary.

\begin{figure*}[!htbp]
 \includegraphics[width=\textwidth]{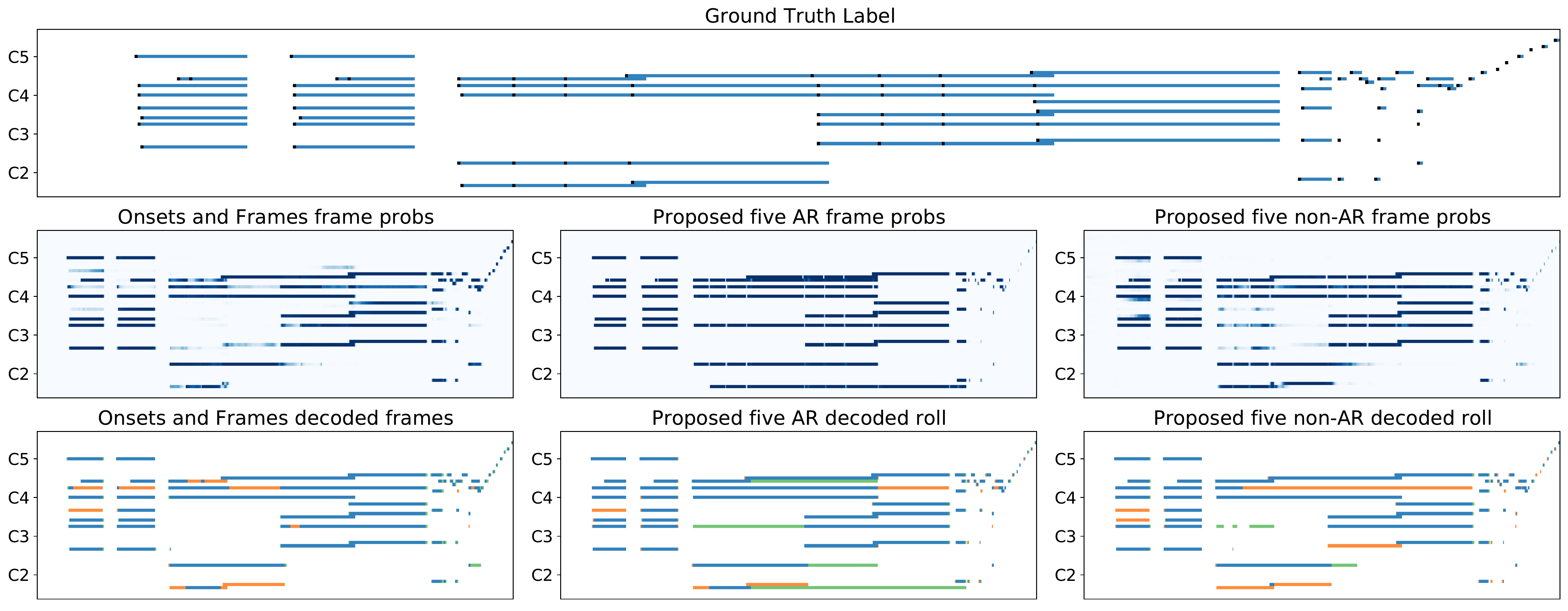}
 \caption{Frame probabilities (or activations) and decoded piano rolls with the ground truth. The excerpt was selected from the test set. \textit{Sustain} frames and \textit{onset} frames are displayed in blue and black, respectively, in the ground truth. In the decoded piano rolls below, the estimated frames with \{true positive, false negative, false positives\} are annotated in blue, orange, and green, respectively. Best viewed in color.}
 \label{fig:states_visualization}
\end{figure*}

\subsection{Effect of Autoregressive Connection}
All AR models with \textit{onset} show similar high performances in the note onset score. While the AR connection improves the F-score in both onset and offset of notes, it significantly decreases the frame-level scores. This might be because some extremely elongated note offsets are predicted by the AR model. This issue is discussed in the following section. The improvement in the note-with-offset score with \textit{re-onset} and \textit{offset} should be carefully understood because our model cannot learn state dependency backward (for example, considering that there will be an offset few frames later, the current frame is more likely to be \textit{on}). Therefore, it is not clear why it is effective only on the AR model. Since our model was trained with a teacher-forcing scenario, we suspect that part of the benefits might be related to resilience on the noisy output, which can occur in the inference phase.


\subsection{Learned State Transition}
\label{sec:state_transition}

\figref{fig:states_visualization} shows an example of frame activations and decoded piano rolls of the selected models. 
In the proposed `Five' model with the AR connection, a note always starts with its onset prediction and the activation is clearly maintained with the following sustain predictions. 
This contrasts with the blurry activation in the \textit{Onsets and Frames} model and the non-AR model, where some notes were estimated too short. 
This is identified by the short blue frames followed by the orange frames in the decoded roll in \figref{fig:states_visualization}. 
Estimating continuous sustain frames also has an negative effect. The AR model often fails to detect note offsets as the sustain frames are estimated too long. This is identified by the green frames followed by the blue frames in the decoded roll in \figref{fig:states_visualization}. 

\begin{table}[!t]
\resizebox{\columnwidth}{!}{%
\begin{tabular}{@{}ccccccc@{}}
\toprule
\multirow{2}{*}{Decoding} & \multirow{2}{*}{F1 score} & \multicolumn{5}{c}{Models}   \\ \cline{3-7} 
                          &  & Five  & Four & Four & Three  & Binary \\
                          &  & & (offset) & (re-onset) & &
                          \\ \midrule
\multirow{2}{*}{Greedy}   & Note Onset & 0.9467 & 0.9416 & 0.9465    & 0.9433 & 0.7560 \\
                          & Note w. Offset          & 0.7936 & 0.7735 & 0.7744    & 0.7593 & 0.5064 \\ \midrule
\multirow{2}{*}{Beam Search}     & Note Onset                & 0.9380 & 0.9305 & 0.9361    & 0.9310 & 0.7480 \\
                          & Note w. Offset          & 0.7886 & 0.7648 & 0.7657    & 0.7484 & 0.5035 \\ \bottomrule
\end{tabular}}

\caption{Summary of note decoding results. All models have autoregressive model.} 
\label{tab:beam_result}
\end{table}

\subsection{Beam Search and Error Calibration}
\label{sec:beam_result}
\begin{figure}[!t]
 \includegraphics[width=\columnwidth]{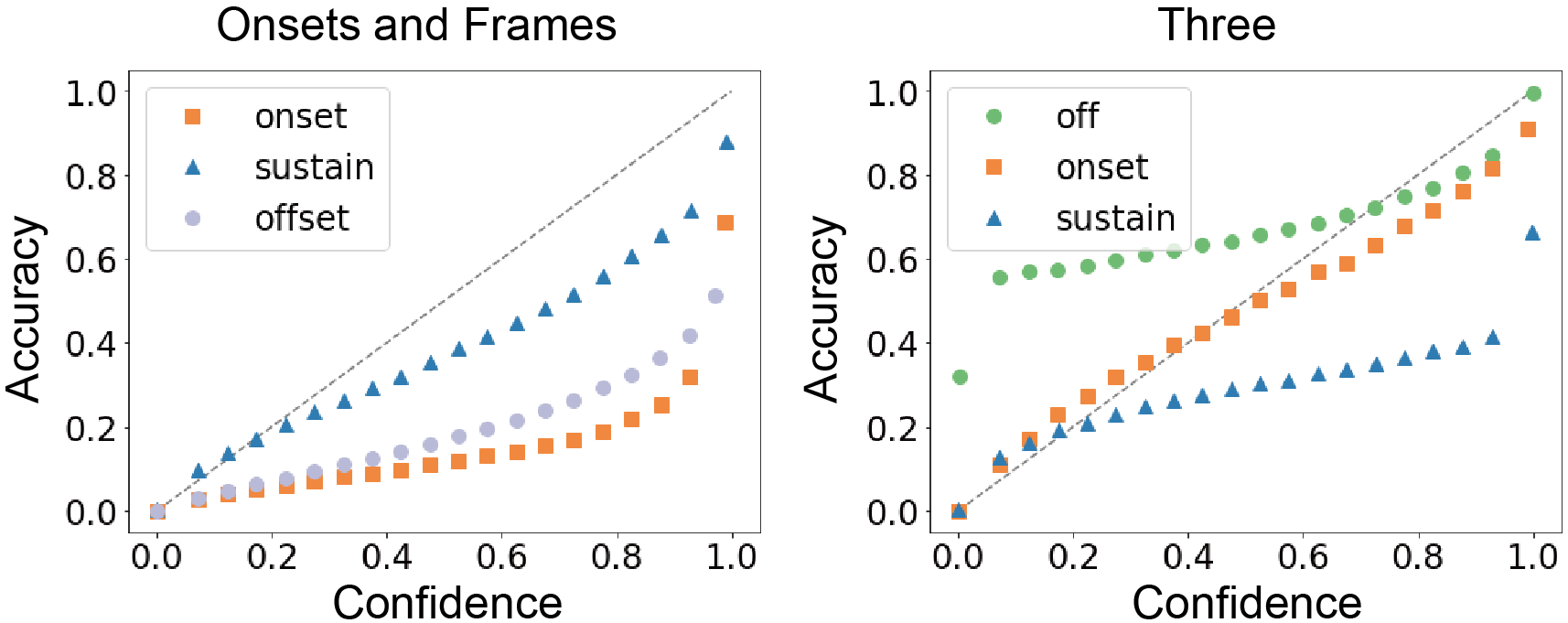}
\caption{Confidence diagrams of the \textit{Onsets and frames} model and the `Three' model. The dashed diagonal line in black indicates perfect calibration. }
 \label{fig:ece}
\end{figure}

We summarized the comparison between the two decoding algorithms in \tablename~\ref{tab:beam_result}. 
Counter-intuitively, the proposed beam search performed slightly worse than the simple greedy decoding method. 
To analyze the reason, we investigated the confidence errors of class prediction. With the greedy decoding, we regarded their softmax predictions as a confidence measure and classified each estimation according to the value. We equally divided the confidence range $[0\ 1]$ into 20 bins and recorded the averaged accuracy of each bin. The empirical discrepancy between accuracy and confidence indicates the model calibration error\cite{guo2017calibration}. The confidence diagram \figref{fig:ece} shows that the model is overconfident on \textit{sustain} prediction but it is under-confident on \textit{off} frames. We suspect this large gap between estimated probabilities leads to sub-optimal paths in the beam search. Label smoothing \cite{muller2019does} or temperature scaling \cite{guo2017calibration} may be helpful for relaxation but we leave this for future work. 


\section{Conclusions}
We proposed a neural network architecture for polyphonic piano transcription where the acoustic and language models are integrated in a unified manner. The architecture is designed to predict multiple note states as a softmax output and learn the dependency among note states through the auto-regressive MLM. Our comparative study shows that the \textit{onset} state is critical to improving note onset scores and the \textit{offset} and \textit{re-onset} states help improving the note-with-offset score. The auto-regressive MLM provides significantly higher accuracy on both note onset and offset estimation compared to its non-auto-regressive version. The visualization of decoded piano roll shows that our models with the auto-regressive connection generates a realistic sequence of note states. We also examined a pitch-wise beam search to decode the frame-level activation but the result showed that it was not as effective as a simple greedy decoding. Finally, the evaluation on the MAESTRO dataset shows that our proposed model achieves transcription performance comparable to the state-of-the-art models even with the unidirectional RNN and fewer parameters. 

\section{Acknowledgement}
This research was funded by Samsung Research Funding
Center under the project number SRFC-IT1702-12

\bibliography{ISMIRtemplate}

%
%
%
%

\end{document}